\documentclass[sigconf]{acmart}
\AtBeginDocument{%
  }

\usepackage{hyperref}
\usepackage{booktabs}
\usepackage{array}
\usepackage{balance} 
\usepackage{multirow}
\usepackage[normalem]{ulem}
\usepackage{color}
\definecolor{lightgray}{RGB}{215,215,215}
\usepackage{colortbl}
\usepackage{xcolor}
\useunder{\uline}{\myul}{}
\usepackage{subfigure}
\usepackage{algorithm}  
\usepackage{algorithmicx}  
\usepackage[noend]{algpseudocode}  
\usepackage{stackengine}

\usepackage{amsmath}  
\usepackage{enumitem}
\usepackage{tabularx}
\usepackage[utf8]{inputenc}
\usepackage[english]{babel}
\usepackage{amsthm}
\usepackage{bm}
\usepackage{threeparttable}

\usepackage[dvipsnames]{xcolor}
\usepackage[most]{tcolorbox}
\usepackage{graphicx}

\newcommand{\ie}{\emph{i.e., }}
\newcommand{\eg}{\emph{e.g., }}

\newcommand{\etc}{\emph{etc.}}

\lstdefinestyle{jsonstyle}{
  basicstyle=\ttfamily\small,
  columns=fullflexible,
  keepspaces=true,
  showstringspaces=false,
  breaklines=true,
  frame=none,
  aboveskip=0pt,
  belowskip=0pt
}

\setcopyright{acmlicensed}
\copyrightyear{2018}
\acmYear{2018}
\acmDOI{XXXXXXX.XXXXXXX}
\acmConference[Conference acronym 'XX]{Make sure to enter the correct
  conference title from your rights confirmation email}{June 03--05,
  2018}{Woodstock, NY}
\acmISBN{978-1-4503-XXXX-X/2018/06}

\begin{document}
\title{NextAds: Towards Next-Generation Personalized Video Advertising}

\author{Yiyan Xu}
\email{yiyanxu24@gmail.com}
\affiliation{
\institution{University of Science and Technology of China}
\country{China}
\city{Hefei}
}

\author{Ruoxuan Xia}
\email{ruoxuanxiaaa@gmail.com}
\affiliation{
\institution{University of Science and Technology of China}
\country{China}
\city{Hefei}
}

\author{Wuqiang Zheng}
\email{qqqqqzheng@gmail.com}
\affiliation{
\institution{University of Science and Technology of China}
\country{China}
\city{Hefei}
}

\author{Fengbin Zhu}
\email{zhfengbin@gmail.com}
\affiliation{
\institution{National University of Singapore}
\country{Singapore}
\city{Singapore}
}

\author{Wenjie Wang}
\email{wenjiewang96@gmail.com}
\affiliation{
\institution{University of Science and Technology of China}
\country{China}
\city{Hefei}
}

\author{Fuli Feng}
\email{fulifeng93@gmail.com}
\affiliation{
\institution{University of Science and Technology of China}
\country{China}
\city{Hefei}
}

\renewcommand{\shortauthors}{Yiyan Xu et al.}

\begin{abstract}
  With the rapid growth of online video consumption, video advertising has become increasingly dominant in the digital advertising landscape. Yet diverse users and viewing contexts make one-size-fits-all ad creatives insufficient for consistent effectiveness, underlining the importance of personalization. In practice, most personalized video advertising systems follow a retrieval-based paradigm, selecting the optimal one from a small set of professionally pre-produced creatives for each user. Such static and finite inventories limit both the granularity and the timeliness of personalization, and prevent the creatives from being continuously refined based on online user feedback. Recent advances in generative AI make it possible to move beyond retrieval toward optimizing video creatives in a continuous space at serving time.
  
  In this light, we propose NextAds, a generation-based paradigm for next-generation personalized video advertising. We articulate its key principles and conceptualize an agentic workflow for closed-loop creative optimization, spanning planning, generation, verification, and reflection. We further explore how personalized creatives can evolve across impressions through feedback-driven refinement and progressive storytelling. To further ground this vision, we formalize two representative tasks and conduct exploratory experiments on corresponding lightweight benchmarks, revealing both the promise and limitations of existing GenAI. Moreover, we discuss potential research opportunities and the broader implications of increasingly personalized and native advertising, highlighting issues that warrant attention from the community and aiming to provide insights for future research and responsible deployment. Our code, benchmarks, and demonstrations are available at \textcolor{magenta}{\url{https://anonymous.4open.science/w/NextAdsProjectPage/}}.
\end{abstract}

\ccsdesc[500]{Information systems~Computational advertising}
\ccsdesc[500]{Information systems~Multimedia content creation}
\ccsdesc[500]{Information systems~Personalization}

\keywords{Video Advertising, Personalized Generation, Agentic Generation}

\maketitle

\section{Introduction}

\begin{figure}[h]
\setlength{\abovecaptionskip}{0.1cm}
\setlength{\belowcaptionskip}{-0.2cm}
\centering
\includegraphics[scale=0.62]{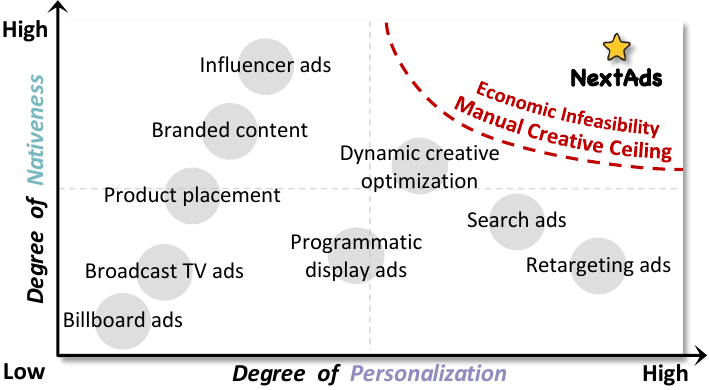}
\caption{The evolution of video advertising has largely followed two dimensions: personalization and nativeness, yet high production costs and manual creative bottlenecks prevent the industry from reaching the upper-right corner.} 
\label{fig:evolution}
\end{figure}

As the financial backbone of the Web ecosystem, digital advertising serves as the primary revenue engine for social media, content platforms, and the broader app economy~\cite{prihatiningsih2025digital}. By funding content creation and delivery at scale, it sustains free Web services while shaping how users access and engage with information. Over the past decades, research and industry practice have largely centered on refining targeting, allocation, and auction mechanisms to expose ads to the right audiences at the right time~\cite{liu2021neural,iyer2005targeting,li2024deep}. As these mechanisms approach saturation, the next frontier of performance gains increasingly hinges on the ad creative itself: whether the ad earns attention, builds trust, and ultimately resonates with its audiences~\cite{yang2024new,chen2021automated}. This elevates creative optimization, which adapts advertising content to specific users and contexts, as a central determinant of advertising effectiveness~\cite{yeo2025persuasive,prihatiningsih2025digital}.

\begin{figure}[t]
\setlength{\abovecaptionskip}{0.1cm}
\setlength{\belowcaptionskip}{-0.3cm}
\centering
\includegraphics[width=\linewidth]{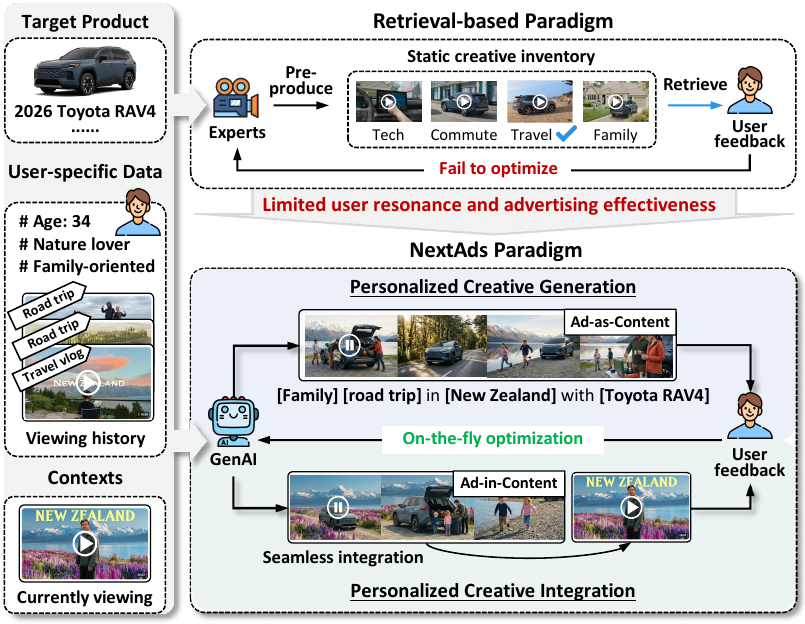}
\caption{\textbf{Paradigm shift: from retrieval to generation.} NextAds transforms personalized video advertising from retrieval over a discrete creative inventory to generative optimization in a continuous space.} 
\Description{A comparison between retrieval-based video advertising and generation-based NextAds optimization.}
\label{fig:intro}
\end{figure}

Among all formats, video stands at the forefront of this creative-centric shift. With user attention increasingly concentrated in video-first platforms such as YouTube and TikTok, video advertising has become one of the fastest-growing segments~\cite{yang2025engagement,zhou2021impact}. Industry forecasts project video advertising to expand by roughly 16\% annually from 2026 to 2033~\cite{GrandViewResearch}. Compared with text and static image ads, video enables richer storytelling and stronger narrative immersion, which can translate into deeper resonance and engagement~\cite{prihatiningsih2025digital}. 
Revisiting the evolution of video advertising through creative optimization, as shown in Fig.~\ref{fig:evolution}, industry practice largely follows two dimensions: \textbf{1) \textit{personalization}}, which measures how well the creative aligns with user preferences and intent; \textbf{2) \textit{nativeness}}, which captures how seamlessly a creative integrates into its surrounding content and viewing context, thereby minimizing intrusiveness. However, video creatives remain expensive, professionally produced artifacts with limited throughput and long production cycles, which restrict the quantity and diversity of creatives. As a result, simultaneously achieving high personalization and high nativeness remains largely unattainable in practice.

Under such constraints, the industry has converged on a retrieval-based paradigm: as shown in Fig.~\ref{fig:intro}, advertisers often employ professional experts to pre-produce a limited set of video creatives or modular assets for the target product, retrieving the most suitable variant to match user preferences and contexts at serving time~\cite{lin2022joint,mishra2020learning}. While practical, this paradigm faces two limitations: \textbf{1) \textit{static inventory constraints}}: such static and finite creative inventories struggle to accommodate rapidly evolving user preferences, thereby limiting user resonance and advertising effectiveness; and \textbf{2) \textit{feedback latency}}: user feedback (\eg dwell time, conversions) only serves to inform subsequent production cycles, failing to refine in-flight video creatives, thus creating a significant latency between performance signals and creative production.

Recent advances in Generative AI (GenAI) offer a promising pathway to transcend these limitations. In particular, 1) modern Vision-Language Models (VLMs), such as GPT~\cite{hurst2024gpt} and Gemini~\cite{comanici2025gemini} have demonstrated exceptional proficiency in instruction-following, multimodal understanding and reasoning, enabling precise user preference modeling for downstream personalization~\cite{xu-etal-2025-personalized,xu2025personalized}. 2) The latest video generation models (\eg Seedance~\cite{seedance2026seedance}, Sora~\cite{sora2_system_card}) facilitate near-instant synthesis of high-fidelity, cinematic video creatives with marginal costs far below manual production. 3) Most importantly, these models support on-the-fly optimization based on online user feedback, enabling creatives to evolve during deployment~\cite{xue2025phyt2v}. Collectively, these advances are driving a paradigm shift from static retrieval to dynamic generation, where video creatives are no longer fixed artifacts selected from a finite pool, but adaptive assets can be synthesized and continuously optimized on-the-fly to better align with individual preferences and contexts. 

In this perspective, we propose \textbf{NextAds}, a generation-based paradigm towards next-generation personalized video advertising. As illustrated in Figure~\ref{fig:intro}, given a target product, user-specific data, and contextual information, NextAds envisions synthesizing personalized creatives that align with individual preferences and contextual constraints while remaining faithful to the underlying product information. To articulate this vision, we further conceptualize an agentic workflow comprising four specialized agents: \textit{\textbf{Director}}, \textbf{\textit{Producer}}, \textbf{\textit{Verifier}}, and \textbf{\textit{Reflector}}, supported by a shared hierarchical memory. These agents collaboratively perform creative planning, generation, verification, and reflection, forming a closed-loop process that enables pre-deployment refinement and, potentially, continuous creative evolution across impressions through online feedback and progressive storytelling.

To illustrate this paradigm in action, we present representative demonstrations covering personalized creative generation, context-aware integration, closed-loop refinement, and cross-impression evolution. Moving beyond, we formalize two representative tasks: \textbf{Personalized Creative Generation (PCG)} and \textbf{Personalized Creative Integration (PCI)}. We further construct two corresponding lightweight benchmarks, \textbf{PCG-Bench} and \textbf{PCI-Bench}, and conduct exploratory quantitative experiments using existing GenAI models. The results demonstrate promising personalization capabilities while revealing key challenges in product alignment, creative diversity, and integration quality. Beyond these initial findings, we identify promising research opportunities for advancing the NextAds paradigm and examine its broader societal implications, particularly regarding privacy, transparency, user autonomy, and accountability. Through these efforts, we aim to establish a starting point for systematic research and encourage the responsible development of next-generation personalized video advertising.

In summary, our contributions are as follows:
\begin{itemize}[leftmargin=*]
    \item \textbf{Paradigm shift.}
    We propose NextAds, shifting personalized video advertising from retrieval over a finite creative inventory to generation-based optimization in a continuous creative space.

    \item \textbf{Conceptual workflow and demonstrations.}
    We conceptualize an agentic workflow for closed-loop creative optimization and present representative demonstrations to illustrate potential instantiations of the NextAds paradigm.

    \item \textbf{Task formulation and empirical analysis.}
    We formalize two representative tasks, PCG and PCI, construct corresponding lightweight benchmarks, conduct exploratory experiments to assess the capabilities and limitations of existing GenAI.

    \item \textbf{Research opportunities and societal implications.}
    We identify key technical challenges and promising research directions, and examine broader societal implications and potential risks to guide future research and responsible deployment.
\end{itemize}

\section{Paradigm Shift: From Retrieval to Generation}

\begin{figure*}[t]
\setlength{\abovecaptionskip}{0.15cm}
\setlength{\belowcaptionskip}{-0.2cm}
\centering
\includegraphics[width=0.98\textwidth]{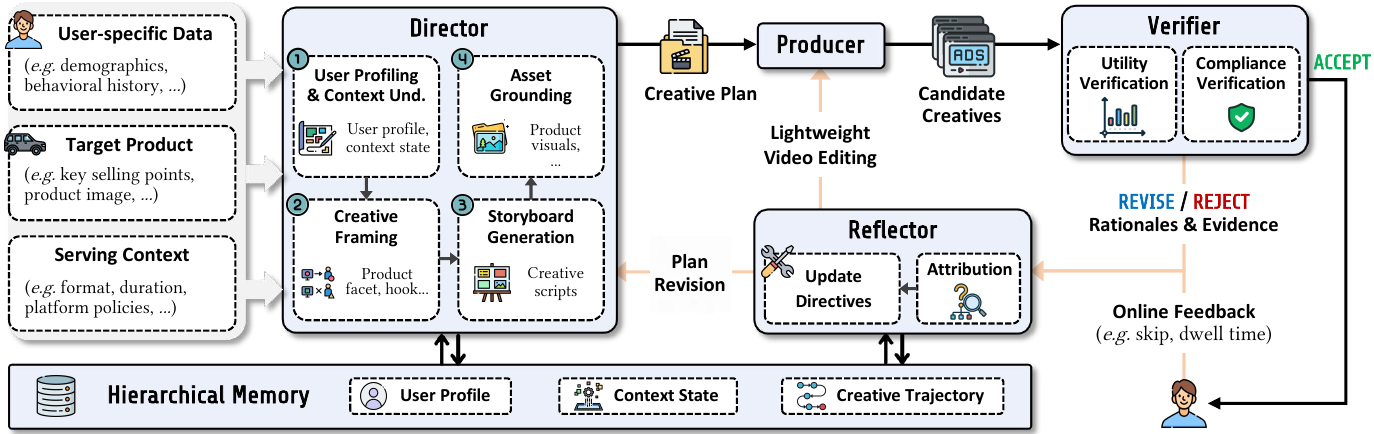}
\caption{\textbf{Agentic workflow of NextAds}, where four core agents, \textit{Director}, \textit{Producer}, \textit{Verifier}, and \textit{Reflector}, collaborate with \textit{Hierarchical Memory} to enable closed-loop refinement through pre-deployment verification and post-deployment user feedback.} 
\label{fig:nextads}
\end{figure*}

Personalized video advertising aims to optimize advertising effectiveness by adapting video ads to individual users and contexts. In this work, we focus on creative optimization, \ie tailoring the video content itself, such as storyline, visuals, and narration.

\vspace{3pt}
\noindent$\bullet$ \textbf{Problem formalization.} Given a user $u$, target product $p$, context $c$, the goal of creative personalization is to produce a video creative $v$ that maximizes the advertising utility:
\begin{equation}
    v = \arg\max_{v\in\mathcal{V}} U(v|u,p,c),
\end{equation}
where $\mathcal{V}$ denotes the creative space, and $U(\cdot)$ is the utility function.

\vspace{3pt}
\noindent$\bullet$ \textbf{Retrieval as a practical approximation.} In practice, optimizing creatives over the continuous creative space $\mathcal{V}$ was impractical due to the high costs, long iteration cycles, and reliance on professional expertise inherent in video production.
As a result, most existing systems adopt a retrieval-based paradigm: advertisers rely on professional teams to pre-produce a finite creative pool $\mathcal{P}=\left\{v_1,\dots,v_N\right\}$, reducing continuous creative optimization to discrete selection:
\begin{equation}
    v = \arg\max_{v\in\mathcal{P}} U(v|u,p,c).
\end{equation}
By approximating the infinite, continuous creative space with a finite discrete creative pool, it induces two inherent limitations: 1) the finite and static creative pool struggles to cover diverse user preferences and context shift; and 2) user feedback can only be incorporated in subsequent production cycles instead of refining in-flight creatives, thereby bounding advertising utility.

\vspace{3pt}
\noindent$\bullet$ \textbf{Revisiting creative personalization with GenAI.} The rapid development of GenAI dramatically changes this landscape, streamlining the video creative production process into a low-cost, low-barrier, and fast-iterating workflow. By effectively lifting production capacity constraints, GenAI enables a return to first-principles optimization over the continuous creative space $\mathcal{V}$.

\section{The NextAds Paradigm}

Here, we present an overview of NextAds, covering its key principles, conceptualized workflow, and closed-loop creative evolution.

\subsection{Key Principles}
NextAds is a generative, closed-loop paradigm for video creative optimization. Empowered by GenAI, it turns creative optimization from \emph{pool-based selection} into \emph{serving-time, closed-loop optimization}. We articulate the paradigm through the following key principles:
\begin{itemize}[leftmargin=*]
    \item \textbf{Specialized collaboration.} Much like professional creative production, video creative optimization is inherently a multi-stage process that requires varied expertise and therefore relies on specialized yet coordinated roles with complementary capabilities.
    \item \textbf{Responsible deployment.} Candidate creatives require rigorous pre-deployment verification along both utility and compliance dimensions, ensuring sufficient effectiveness while satisfying strict constraints such as platform policies and privacy boundaries.
    \item \textbf{Continuous closed-loop optimization.} Rather than selecting creatives from a pre-produced video pool, NextAds continually synthesizes and refines creatives at serving time, using verification outcomes and online feedback for timely optimization.
\end{itemize}

\subsection{Agentic Workflow}
Motivated by the above principles, as depicted in Fig.~\ref{fig:nextads}, we conceptualize NextAds as an agentic workflow with a shared memory module and four core agents, decomposing creative optimization into planning, generation, verification, and reflection.

\subsubsection{\textbf{Hierarchical Memory}}
NextAds maintains a \textit{Hierarchical Memory} consisting of three types of states accumulated throughout creative optimization: user profile, context state, and creative trajectory. User profile stores reusable user-specific information, including topical interests (\eg gaming, travel), preferred presentation styles (\eg humorous, informative), implicit needs (\eg quality-of-life, self-expression), and decision-making tendencies (\eg price-sensitive, quality-oriented), which can be reused across creatives. Context state captures the current serving situation, including surrounding content, delivery constraints, and platform requirements. Creative trajectory records previously delivered creative decisions, including hooks, narrative angles, verification outcomes, and observed user feedback. These states are constructed and updated by different components throughout the workflow.

\subsubsection{\textbf{Director}}
As the strategic planning module, \textit{Director} is responsible for generating a personalized and context-aware creative plan. We break down the procedure into four steps:
\begin{itemize}[leftmargin=*]
    \item \textbf{User profiling and context understanding.} \textit{Director} first retrieves the existing user profile and incrementally updates it with newly observed user-specific data. Meanwhile, it refreshes the context state based on the current serving situation. The updated states are stored back into \textit{Hierarchical Memory} to guide subsequent creative planning.

    \item \textbf{Creative framing.} Given the product information and \textit{Hierarchical Memory}, \textit{Director} determines how the current creative should frame the product. It prioritizes user- and context-relevant product facets, selects suitable topics or scenarios to contextualize them, and defines the core narrative angle. The creative trajectory in \textit{Hierarchical Memory} further informs this decision, helping maintain continuity across impressions while avoiding redundant messages and inconsistent narrative progression.

    \item \textbf{Storyboard generation.} Subsequently, \textit{Director} produces an executable, scene-level storyboard script that specifies the full ad content, namely what appears in each scene (\eg key messages, visuals) and how scenes are ordered and paced, ensuring a coherent flow from attention capture to the final call to action (CTA) under the serving contexts.

    \item \textbf{Asset grounding.} To support faithful generation, \textit{Director} binds storyboard elements to eligible product assets (\eg product images, logos) and user-preferred visual cues, producing grounding constraints that guide subsequent generation and verification.
\end{itemize}
Overall, given user-specific data, a target product, and serving contexts, \textit{Director} constructs a structured personalized creative plan with an executable storyboard and grounded assets, serving as a controllable blueprint for subsequent generation and verification.

\subsubsection{\textbf{Producer}} As the execution module, \textit{Producer} translates the high-level creative plan into candidate creatives. Given a creative plan, including the storyboard, grounded assets, and generation constraints, \textit{Producer} selects and invokes appropriate tools to accomplish the required operations. These operations may include full video synthesis, localized video editing, partial segment re-generation, caption insertion, and voiceover synthesis.

\subsubsection{\textbf{Verifier}} 
\label{sec:verifier}
Given a candidate creative, its creative plan, and supporting evidence, \textit{Verifier} conducts two assessments:
\begin{itemize}[leftmargin=*]
    \item \textbf{Utility verification.} It estimates the quality and effectiveness: 1) \textit{video quality}, including audiovisual clarity and temporal stability; 2) \textit{personalization}, assessing its appeal to the target user; 3) \textit{product alignment}, examining consistency with the target product information; and 4) \textit{context alignment}, assessing its fit with the surrounding content and delivery constraints.

    \item \textbf{Compliance verification.} It determines the eligibility for deployment: 1) \textit{safety}, including unsafe topics, deceptive patterns, and platform policy violations; 2) \textit{privacy and personalization boundaries}, concerning the inference, disclosure, or implication of sensitive personal information; 3) \textit{fairness and bias}, covering discriminatory or stereotypical language and visuals; and 4) \textit{copyright, portrait rights, and asset licensing}, concerning the authorization and licensing of visuals, voices, music, identifiable individuals, and derivative content.
\end{itemize}

Creatives that pass all checks are marked accepted and prepared for online deployment with required disclosures, such as \emph{AI-Created} labels. Creatives with severe utility failures or compliance violations are rejected, while those with minor or localized issues are marked for revision. Along with each decision, \textit{Verifier} provides structured rationales and localized evidence, such as the triggering claim or visual segment, to support traceable diagnosis and subsequent reflection and refinement.

\subsubsection{\textbf{Reflector}}
As the feedback-to-refine engine, \textit{Reflector} converts pre-deployment assessments and post-deployment interaction signals into actionable update directives. It receives two sources of feedback: 1) verification results from \textit{Verifier}, such as low utility scores, problematic visual segments, or compliance risks, which provide explicit diagnostic evidence; and 2) online user feedback, such as dwell time and conversions, which provides implicit preference evidence. Based on these, \textit{Reflector} performs two key functions:
\begin{itemize}[leftmargin=*]
    \item \textbf{Attribution.} \textit{Reflector} attributes feedback to specific creative decisions. Verification feedback is typically localized and evidence-driven, identifying plan-level issues (\eg weak user-product matching or unsafe claims) or execution-level issues (\eg visual artifacts or inconsistencies with grounded assets). For online feedback, positive signals reinforce effective creative choices, while negative signals may indicate weak resonance, creative fatigue, or poor contextual fit.

    \item \textbf{Directive planning.} Based on the attribution result, \textit{Reflector} produces minimal update directives for refinement. Verification failures trigger targeted revisions, such as storyboard revision and localized editing. Positive online feedback reinforces effective hooks or visual styles, while negative feedback prompts adjustments to creative angles or contextual fit. When feedback is sufficiently reliable, \textit{Reflector} also updates the user profile or creative trajectory to preserve effective strategies, reduce repetition, and improve future personalization.
\end{itemize}

\begin{figure*}[t]
\setlength{\abovecaptionskip}{0.1cm}
\setlength{\belowcaptionskip}{-0.2cm}
\centering
\includegraphics[width=\linewidth]{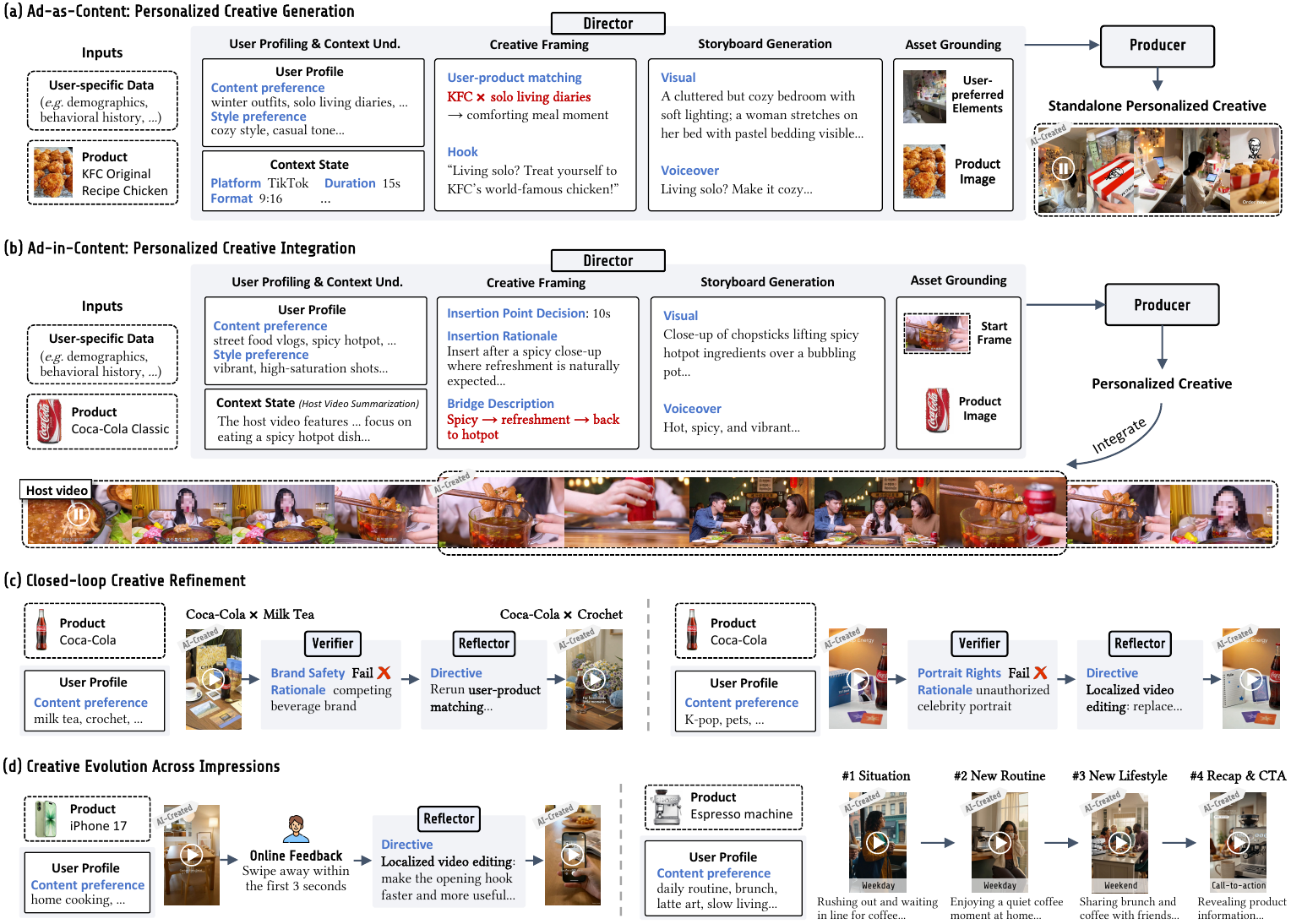}
\caption{\textbf{NextAds in action:} (a) For standalone creative generation, NextAds transforms user history and product information into a personalized creative plan and realizes it as a video ad. (b) For context-aware creative integration, NextAds additionally conditions on the host video to identify a natural insertion point and generate an ad segment that blends into the viewing context. Beyond one-shot generation, Verifier and Reflector close the loop by (c) revising problematic creatives before deployment and (d) evolving creatives across impressions through feedback-driven refinement and progressive storytelling.}
\label{fig:demo}
\end{figure*}

\subsection{Creative Evolution Across Impressions}
Advertising is inherently temporal, with campaigns repeatedly reaching the same user across multiple impressions. Under the traditional retrieval-based paradigm, creatives are produced offline and remain largely static, often replaying the same creative or a small set of variants. This limits adaptation to new user feedback, context shifts, and creative fatigue. NextAds instead enables \textbf{serving-time, closed-loop creative evolution} across impressions.

\subsubsection{\textbf{Iterative Refinement.}}
Across repeated impressions, NextAds adapts subsequent creatives to new user feedback and evolving serving contexts through: 1) \textit{feedback-driven refinement}, adjusting hooks, pacing, tone, or presentation based on user reactions; and 2) \textit{context-responsive adaptation}, adjusting creative presentation to changes in viewing content, time, or delivery format.

\subsubsection{\textbf{Progressive Storytelling.}}
Beyond refining individual creatives, NextAds can optimize the broader creative trajectory through progressive storytelling. Unlike iterative refinement, later impressions build on earlier ones, with each serving a distinct role in advancing the user--product narrative rather than repeatedly optimizing the same message. Specifically, early impressions may establish a user-relevant need or lightweight product association, subsequent ones introduce complementary use cases or product values, and later ones strengthen purchase considerations and CTAs. The system therefore decides not only \emph{how to improve the next creative}, but also \emph{what to communicate next}. Guided by the creative trajectory in \textit{Hierarchical Memory}, \textit{Director} can reinforce the current narrative, extend it with complementary angles, or pivot when needed, while reducing repetition and creative fatigue.

\section{Demonstrations}

To illustrate NextAds in action, Fig.~\ref{fig:demo} presents four representative cases, focusing on the feasibility and potential formats of the paradigm.
This case study is mainly based on GPT-4o and Sora2, with user information drawn from Qilin~\cite{qilin} and MicroLens~\cite{microlens} datasets, supplemented by a few LLM-simulated users. Due to space constraints, we provide only the essential parts here, with more cases and details available at our project page\footnote{\url{https://anonymous.4open.science/w/NextAdsProjectPage/}.}. 

\textbf{Ad-as-Content: Personalized Creative Generation (PCG).}
It targets standalone video ads, such as those interleaved with organic posts in TikTok-style feeds\footnote{\url{https://ads.tiktok.com/business/creativecenter/tiktok-topads-spotlight/}.}, where the creative itself must capture attention and communicate the product message. As shown in Fig.~\ref{fig:demo}(a), NextAds infers user preferences from behavioral history and aligns them with the target product. For KFC Original Recipe Chicken, \textit{Director} selects \textit{solo living diaries} as a compatible narrative angle, develops a storyboard and grounds it with user-preferred visual elements and official product assets, which \textit{Producer} then realizes as a personalized creative.

\textbf{Ad-in-Content: Personalized Creative Integration (PCI).}
It targets ads embedded in consumed content, such as personalized segments inserted into long videos on platforms like YouTube\footnote{\url{https://business.google.com/en-all/ad-solutions/youtube-ads/best-ads/}.}. Rather than relying on generic pre-roll or mid-roll ads, NextAds generates personalized segments coherent with the surrounding content. As shown in Fig.~\ref{fig:demo}(b), for a host video featuring spicy hotpot, \textit{Director} identifies a natural insertion point and constructs a bridge from \textit{spicy food} to \textit{refreshment} and back to the host content. Grounded by the selected start frame and product asset, \textit{Producer} realizes the plan as a seamlessly integrated personalized segment.

\textbf{Closed-loop Creative Refinement.}
Fig.~\ref{fig:demo}(c) illustrates closed-loop refinement before deployment. \textit{Verifier} localizes problematic content and provides failure rationales, which \textit{Reflector} attributes to either planning or execution. In the first case, a Coca-Cola creative framed around milk tea introduces a competing beverage brand, prompting \textit{Reflector} to rerun user--product matching and switch to a compatible crochet scenario. In the second, an unauthorized celebrity portrait triggers a portrait-rights violation, so \textit{Reflector} directs \textit{Producer} to replace the problematic visual.

\textbf{Creative Evolution Across Impressions.}
Fig.~\ref{fig:demo}(d) extends the closed loop to post-deployment feedback. For example, an early swipe signals a weak opening and prompts \textit{Reflector} to make the next creative faster and more informative. Over longer horizons, NextAds can coordinate impressions into a coherent narrative that progressively builds user need and product relevance. In the espresso-machine example, the story first establishes a familiar weekday pain point of rushing out and waiting for coffee, then introduces a cozy at-home coffee routine, expands it into a weekend brunch-and-coffee lifestyle, and finally consolidates the narrative with product information and a stronger CTA.

\section{Benchmarks and Experiments}
To move beyond the conceptual paradigm and facilitate systematic research, we formalize PCG and PCI as two reproducible tasks, construct corresponding benchmarks, and conduct preliminary experiments to investigate the capabilities of current GenAI.

\subsection{Task Formulation}
\subsubsection{\textbf{PCG}} 
This task aims to produce standalone personalized video ads that are consumed independently as content.
\begin{itemize}[leftmargin=*]
    \item \textbf{Inputs:} 1) User-specific data, such as demographics and behavioral history; 2) target product, including key selling points, and eligible assets like product images and logo; and 3) serving context, such as aspect ratio and platform policies.
    \item \textbf{Outputs:} 1) Detailed creative plan as an intermediate artifact; and 2) personalized video creative.
    \item \textbf{Objectives and constraints:} As specified in Section \ref{sec:verifier}, PCG aims to maximize advertising utility along three dimensions: video quality, personalization, and product alignment. In addition, the creative must pass the compliance verification and required ``AI-Created'' disclosure.
\end{itemize}

\subsubsection{\textbf{PCI}} This task aims to seamlessly integrate advertising messages into an existing host video in a personalized manner without degrading the viewing experiences. 
\begin{itemize}[leftmargin=*]
    \item \textbf{Inputs:} Same as the PCG task (user-specific data, target product, and serving context), plus a host video to be integrated with.
    \item \textbf{Outputs:} 1) Detailed creative plan as an intermediate artifact; and 2) personalized video creative with the integration point.
    \item \textbf{Objectives and constraints:} PCI optimizes advertising utility while preserving the host content experience. In addition to video quality, personalization, and product alignment, PCI emphasizes integration quality, requiring low intrusiveness to the host video. Similarly, the creative must pass the compliance verification and required ``AI-Created'' disclosure.
\end{itemize}

\begin{table*}[t]
\setlength{\abovecaptionskip}{0.1cm}
\setlength{\belowcaptionskip}{0.1cm}
\caption{Overall performance comparison of existing GenAI in both PCG and PCI tasks. The best results are highlighted in bold.}
\resizebox{\textwidth}{!}{
\begin{tabular}{l|cc|c|c|c|cccc}
\hline
\textbf{\#PCG-Bench}                & \multicolumn{2}{c|}{\textbf{Personalization}} & \multirow{2}{*}{\textbf{Product Alignment}} & \multirow{2}{*}{\textbf{Integration Quality}} & \multirow{2}{*}{\textbf{Diversity}} & \multicolumn{4}{c}{\textbf{Video Quality}}                                                                    \\ \cline{2-3} \cline{7-10} 
\multicolumn{1}{c|}{\textbf{}}      & \textbf{Presentation}    & \textbf{Content}   &                                             &                                               &                                     & \textbf{MS} & \textbf{DD} & \textbf{AQ} & \textbf{IQ} \\ \hline
\textbf{GenericAds (Closed)} & 1.77                     & 1.52               & 8.48                                        & -                                             & \textbf{7.10}                       & \textbf{0.9756}            & 0.9742                  & \textbf{0.5494}             & \textbf{0.6524}          \\
\textbf{NextAds (Closed)}   & \textbf{7.12}            & \textbf{5.09}      & \textbf{8.58}                               & -                                             & 6.57                                & 0.9406                     & 0.9797                  & 0.5259                      & 0.6339                   \\
\textbf{NextAds (Open)}     & 6.01                     & 3.35               & 3.44                                        & -                                             & 5.40                                & 0.9693                     & \textbf{0.9800}         & 0.4064                      & 0.6199                   \\ \hline
\end{tabular}}
\vspace{2mm}
\resizebox{\textwidth}{!}{
\begin{tabular}{l|cc|c|c|c|cccc}
\hline
\textbf{\#PCI-Bench}                & \multicolumn{2}{c|}{\textbf{Personalization}} & \multirow{2}{*}{\textbf{Product Alignment}} & \multirow{2}{*}{\textbf{Integration Quality}} & \multirow{2}{*}{\textbf{Diversity}} & \multicolumn{4}{c}{\textbf{Video Quality}}                                                                    \\ \cline{2-3} \cline{7-10} 
\multicolumn{1}{c|}{\textbf{}}      & \textbf{Presentation}    & \textbf{Content}   &                                             &                                               &                                     & \textbf{MS} & \textbf{DD} & \textbf{AQ} & \textbf{IQ} \\ \hline
\textbf{GenericAds (Closed)} & 3.97                     & 2.24               & \textbf{9.58}                               & 3.89                                          & \textbf{9.64}                       & \textbf{0.9873}            & \textbf{0.9375}         & 0.5082                      & 0.6422                   \\
\textbf{NextAds (Closed)}   & \textbf{8.56}            & \textbf{8.32}      & 7.89                                        & \textbf{8.67}                                 & 2.80                                & 0.9867                     & 0.9360                  & \textbf{0.5243}             & \textbf{0.6508}          \\
\textbf{NextAds (Open)}     & 5.44                     & 4.20               & 6.26                                        & 6.27                                          & 7.58                                & 0.9771                     & 0.9320                  & 0.4385                      & 0.5986                   \\ \hline
\end{tabular}
}
\label{table:main_results}
\vspace{-0.3cm}
\end{table*}

\subsection{Benchmark Construction}
We construct two benchmarks for the PCG and PCI tasks, where each benchmark includes $50$ users. For each user, we select five target products, yielding $250$ user-product pairs per benchmark.

\vspace{3pt}
\textbf{Product library.} We first build a product library consisting of $18$ products spanning electronics (\eg smartphones), retail goods (\eg shoes, skincare, and food), and platform services (\eg streaming and ride-hailing). For each product, we collect a bundle of eligible creative assets, including product images, brand logos, slogans, and detailed descriptions. This product library serves as a unified source of candidate assets for benchmark construction and evaluation.

\subsubsection{\textbf{PCG-Bench}} 
Built on the Qilin recommendation dataset\footnote{\url{https://github.com/RED-Search/Qilin}.}~\cite{qilin}, where each user is associated with a historical interaction sequence of multimodal notes with metadata (title, image, and text), we select $50$ representative users with distinctive and consistent preferences, ensuring their histories provide sufficient signals for user modeling. For each selected user, we pair the five most compatible products from the product library to form user-product pairs.

\subsubsection{\textbf{PCI-Bench}} 
Built on the MicroLens dataset\footnote{\url{https://github.com/westlake-repl/MicroLens}.}~\cite{microlens}, a large-scale micro-video recommendation dataset that provides user interaction histories and video information (titles, cover images, and full-length videos), we similarly select 50 representative users with clear preferences and pair each with five best-matched products from the product library. For each user sequence, we designate the last interacted micro-video as the \emph{host video} for creative integration, and use the remaining videos as user-specific data for user modeling. This setup reflects a realistic advertising scenario in which the system leverages prior behaviors to personalize current creative.

\subsubsection{\textbf{Evaluation Suites}} 
To comprehensively evaluate the generated creatives, we design a multi-aspect evaluation suite covering video quality, personalization, product alignment, integration quality, and diversity. For \textbf{video quality}, we adopt Motion Smoothness (MS), Dynamic Degree (DD), Aesthetics Quality (AQ), and Imaging Quality (IQ), as suggested in VBench~\cite{huang2024vbench} to assess basic video fidelity. 
The remaining aspects require multimodal understanding and reasoning, which are hard to capture with conventional quantitative metrics, while human evaluation is costly and difficult to reproduce. Therefore, we leverage Gemini~2.5-Flash~\cite{comanici2025gemini} as an automatic judge: it scores each criterion on a $0$-$10$ scale based on predefined rules and simultaneously provides detailed rationales to support its judgments. Specifically, for \textbf{personalization}, we present the user’s historically interacted content alongside the generated creative to assess alignment in presentation (\eg tone, artistic style) and content (\eg topics, scenes, appeal). For \textbf{product alignment}, we provide product information with the creative to evaluate product identity, brand elements, selling points, and factual consistency. For \textbf{diversity}, we assess similarity among all creatives generated for the same user in visual style, narrative structure, topics, and scenes. For \textbf{integration quality} in the PCI task, we insert the generated creative into the corresponding host video and evaluate contextual flow, thematic consistency, and user experience disruption. The evaluation prompts can be found at Appendix~\ref{appendix:eval}.

\subsection{Results}
To investigate the feasibility of instantiating NextAds, we conduct exploratory experiments on both the PCG and PCI tasks. We implement \textit{Director} and \textit{Producer} using both open-source and closed-source GenAI: 1) \textbf{NextAds (open-source)}: Qwen3-VL\footnote{\url{https://huggingface.co/Qwen/Qwen3-VL-8B-Instruct}.} as \textit{Director} and Wan2.2\footnote{\url{https://huggingface.co/Wan-AI/Wan2.2-TI2V-5B}.} as \textit{Producer}; and 2) \textbf{NextAds (closed-source)}: GPT-4o~\cite{hurst2024gpt} as \textit{Director} and Sora2~\cite{sora2_system_card} as \textit{Producer}. We also include \textbf{GenericAds (closed-source)}, where \textit{Director} conditions only on the target product and serving context. It uses no user-specific data in either task; for PCI, it also excludes the host video.

As shown in Table~\ref{table:main_results}, NextAds consistently improves personalization over GenericAds on both PCG and PCI while maintaining comparable video quality, demonstrating the potential of current GenAI models to exploit user-specific signals for personalized creative generation. On PCG, this improvement is achieved with relatively minor changes in product alignment, although the slight decrease in diversity suggests a trade-off between stronger personalization and creative variation. On PCI, the gains are more pronounced: in addition to substantially improved personalization, NextAds achieves markedly better integration quality by explicitly conditioning on the host video and selecting contextually appropriate insertion points. This more constrained setting, however, makes product grounding more challenging and further narrows the feasible generation space, resulting in lower product alignment and diversity. These results highlight a shared challenge across both tasks: improving personalization while preserving product fidelity and creative diversity, with PCI additionally requiring seamless integration into the surrounding content.

Comparing different model configurations reveals a similar pattern across the two tasks. Closed-source NextAds generally achieves stronger personalization and product alignment, suggesting better capability in multimodal understanding, complex conditioning, and faithful execution of detailed storyboard and grounding constraints. In contrast, the open-source variant can achieve comparable video quality, particularly on PCG, but remains weaker in personalization and product alignment. Interestingly, it sometimes attains higher diversity, especially on PCI; rather than indicating stronger creative capability, this likely reflects weaker controllability and instruction following, causing the model to adhere less strictly to storyboard, grounding, and integration constraints. Conversely, stronger conditioning in closed-source models improves alignment and controllability but can further restrict the feasible generation space, exacerbating diversity collapse. This points to an important direction for future work: improving controllability and conditional fidelity without sacrificing creative diversity.

\subsection{Discussion}
During the experiments, we observe two representative failure modes: 1) \textbf{spurious conditioning}: generative models may over-interpret high-level user preferences as explicit visual cues, introducing irrelevant or conflicting elements. For example, an Adidas creative targeting a Nike-interested user may inadvertently incorporate Swoosh-like patterns, compromising product fidelity. 2) \textbf{forced personalization}: when user preferences poorly align with the target product, forcing personalization may produce incoherent creatives, such as placing an industrial-style razor in a soft pastel setting. These observations highlight the importance of selective preference modeling, stronger product grounding, and fallback strategies when user--product compatibility is low.

We also identify two practical insights for effective instantiation: 1) \textbf{decoupling planning from execution} separates \textit{Director} from \textit{Producer}, producing interpretable creative plans that facilitate verification, targeted revision, error attribution, and modular upgrades. 2) \textbf{multi-source visual grounding} leverages complementary references, including target product images, user-preferred visual elements, and host-video insertion frames, to better preserve product fidelity, personalization, and contextual coherence. Together, these design choices offer practical guidance for building more controllable and faithful personalized video advertising.

\section{Potential Research Opportunities}
In this section, we highlight several promising research directions.

\vspace{3pt}
\noindent $\bullet$ \textbf{Evaluation.} 
Although candidate creatives can be evaluated before deployment, their actual utility is more reliably assessed through online user feedback. In NextAds, the \textit{Verifier} performs offline evaluation: generic creative quality can be assessed using established video-generation metrics such as VBench~\cite{huang2024vbench}, while personalization, product alignment, and context alignment can be evaluated via LLM-as-a-Judge with well-designed rubrics. However, such offline metrics remain imperfect proxies for online utility. A promising direction is to develop realistic user simulators or reward models~\cite{usersimulator1, usersimulator2} that approximate user responses, supporting: 1) \textit{pre-deployment utility estimation} within the \textit{Verifier}; and 2) \textit{offline iterative optimization}, by providing fine-grained feedback for repeatedly evaluating and refining creatives without costly online experimentation.

\vspace{3pt}
\noindent $\bullet$ \textbf{Continual user modeling.}
NextAds requires accurate and continually updated user profiles to guide personalized creative generation. This is challenging because advertising feedback (\eg behavioral history and conversions) is largely implicit, noisy, and confounded by exposure contexts such as time and location. Moreover, user preferences are non-stationary, requiring models to capture transient intent while preserving stable long-term interests. Promising directions include: 1) causal and debiasing methods to disentangle genuine preferences from exposure effects~\cite{causalrec1, causalrec2}; and 2) well-designed memory mechanisms that organize preferences across multiple timescales for longitudinal personalization~\cite{memoryrec1, memoryrec2}. 
Meanwhile, increasingly fine-grained user modeling also raises greater data privacy concerns. Federated learning~\cite{federatedlearningpersona} and other privacy-preserving techniques~\cite{privacypersona} offer promising directions for limiting unnecessary exposure of user information.

\vspace{3pt}
\noindent $\bullet$ \textbf{Quality and latency.} 
While GenAI substantially expands the creative space, its computational cost and generation latency pose challenges to latency-sensitive ad serving. Balancing high-quality generation with efficient serving therefore requires mechanisms that reduce or selectively allocate generation overhead, including: 1) \textit{predictive pre-generation}, which anticipates likely user--product matches and pre-generates candidate creatives before serving; 2) \textit{streaming generation}, which serves a pre-generated opening segment while progressively synthesizing subsequent content conditioned on real-time user feedback; and 3) \textit{ROI-aware generation}, which invokes expensive generation only when its expected utility gain outweighs the additional computational cost. By estimating the expected uplift of personalized generation over retrieved creatives, the system can dynamically allocate generation resources to creatives with the highest marginal value.

\vspace{3pt}
\noindent $\bullet$ \textbf{Personalization and diversity.} 
NextAds may carry the risk of over-optimizing towards user preferences, potentially leading to filter bubbles and creative fatigue. To mitigate this, one promising approach is to prioritize exploration strategies within the generative process~\cite{liu2023personalized}. Rather than relying exclusively on exploiting known preferences, controlled stochasticity is introduced to inject novel narrative styles, visual aesthetics, or unexpected hooks. 
This mechanism offers two key functions:
1) \textit{fatigue mitigation}, which circumvents the diminishing returns of redundant ad creatives by maintaining high informational entropy. 
2) \textit{continual user preference refinement}, which enables a broader range of user feedback collection from diverse creatives, thereby enriching the understanding of the user’s evolving preference boundaries.

\vspace{3pt}
\noindent $\bullet$ \textbf{Synergy between targeting, auction, and creative optimization.} 
NextAds fundamentally shifts personalized video advertising from a static, retrieval-based selection task to a dynamic, generation-based optimization problem. Traditional targeting mechanism acts largely as a filter, matching a fixed, pre-produced creative to the most relevant audience. By contrast, NextAds concentrates targeting on identifying the highest-value user–product pairs, then uses GenAI to synthesize personalized creatives that connect them. To make this paradigm practical, two research directions stand out: 1) \textit{augment targeting with generatability prediction.} Beyond estimating conversion propensity, targeting should predict whether personalization is feasible for a given user–product pair, \ie the expected headroom of generation under available assets and contextual constraints. 2) \textit{design auctions that trade off generation cost and expected lift.} For each user–product pair, the mechanism should jointly optimize targeting and generation to decide whether to serve a pre-produced creative or invest in a more expensive personalized one, maximizing expected net value.

\section{Societal Implications and Associated Risks}
As such paradigms become increasingly technically feasible, several broader implications deserve sustained attention.

\vspace{3pt}
\noindent $\bullet$ \textbf{Privacy.}
Fine-grained user modeling may infer latent preferences, needs, or sensitive attributes that users have never explicitly disclosed. Beyond data privacy itself, this raises a broader question of what user information should be allowed to influence advertising and where appropriate personalization boundaries should lie.

\vspace{3pt}
\noindent $\bullet$ \textbf{Transparency.}
Native creative integration can reduce disruption by improving coherence with surrounding content. Yet greater nativeness also blur the boundary between organic content and commercial persuasion, making it harder for users to recognize when content is advertising, AI-created, and personalized for them.

\vspace{3pt}
\noindent $\bullet$ \textbf{User autonomy.}
Closed-loop optimization can progressively learn which creative strategies are most effective for an individual from behavioral feedback. While this may improve relevance, it may also enable increasingly individualized persuasion, raising concerns about the boundary between effective personalization and undue influence on user decisions.

\vspace{3pt}
\noindent $\bullet$ \textbf{Accountability.}
AI-created creatives vary across users and impressions, complicating large-scale oversight and the traceability of problematic content, thereby posing challenges to accountability and systematic auditing in digital information systems.

\section{Related Work}
\noindent $\bullet$ \textbf{Video Advertising.}
As depicted in Fig.~\ref{fig:evolution}, the advertising landscape can be characterized along two dimensions: \textit{personalization} and \textit{nativeness}~\cite{qian2025vcllmautomatedadvertisementvideo, videoad, digitaladsurvey}. Traditional formats, such as billboards and broadcast TV ads, are limited on both dimensions, relying on static creatives and mass delivery. Targeted advertising, such as programmatic display, search, and retargeting ads, improves personalization through audience or intent matching, while creatives remain largely generic and weakly coupled with the viewing context. Conversely, native formats, such as product placement, branded content, and influencer advertising, achieve stronger nativeness but are difficult to personalize at scale. Dynamic creative optimization partially narrows this gap, yet remains constrained by finite assets and costly production workflows. These constraints create a practical frontier that makes simultaneously achieving high personalization and high nativeness difficult. By leveraging GenAI for scalable, serving-time creative generation, NextAds aims to push beyond this frontier toward more personalized and native video advertising.

\vspace{3pt}
\noindent $\bullet$ \textbf{Generative AI for Advertising.}
Recent advances in GenAI have enabled scalable generation of advertising creatives. Existing studies leverage Diffusion Models (DMs) and Large Language Models (LLMs) to generate personalized ad images and banners~\cite{vashishtha2024chaining}, and further incorporate engagement objectives such as click-through rate to optimize creative utility~\cite{chen2025ctr}. More broadly, agentic multimodal frameworks have been explored to automate personalized advertising workflows~\cite{sakhinana2025agentic}. Nevertheless, existing efforts remain predominantly focused on static image creatives, while video advertising has received limited systematic attention. The rapid progress of video generation models therefore creates an opportunity to revisit video advertising from a generation-first perspective.

\vspace{3pt}
\noindent $\bullet$ \textbf{Personalized Generation.} 
Large generative models, including DMs, LLMs, and VLMs, have stimulated growing interest in personalized generation across modalities~\cite{xu-etal-2025-personalized}. By modeling individual preferences and conditioning generation accordingly, these methods have demonstrated strong capabilities in producing personalized content~\cite{qiu-etal-2025-measuring,xu2024diffusion,xu2025drc}. Nevertheless, personalized ad creative generation remains underexplored, and limited efforts mostly focus on images~\cite{yang2024new,shilova2023adbooster}. In contrast, personalized video creative generation, with strong storytelling capacity, is still lacking systematic investigation. This gap motivates our work, which takes a first step toward personalized video advertising by reframing it from retrieval over a static inventory to generation-based creative optimization.
\section{Conclusion}
In this work, we envisioned \textbf{NextAds}, a generation-based paradigm that shifts personalized video advertising from retrieval over a finite creative inventory to serving-time, closed-loop optimization in a continuous creative space. We conceptualized an agentic workflow for personalized planning, grounded generation, utility-and-compliance verification, and feedback-driven refinement, and presented representative demonstrations to illustrate its potential. To further ground this vision, we formalized two representative tasks and conducted exploratory experiments on two corresponding lightweight benchmarks, highlighting both the promise and limitations of existing GenAI. Building on these insights, we outlined key research challenges and opportunities for realizing this paradigm. Looking ahead, we hope NextAds provides a foundation for advancing generative personalized advertising while fostering critical discussions on its societal implications, safeguards, and governance requirements for responsible deployment.


\bibliographystyle{ACM-Reference-Format}
\bibliography{bibfile}

\appendix

\begin{figure*}[t]
\setlength{\abovecaptionskip}{0.2cm}
\setlength{\belowcaptionskip}{-0.2cm}
\centering
\includegraphics[width=\linewidth]{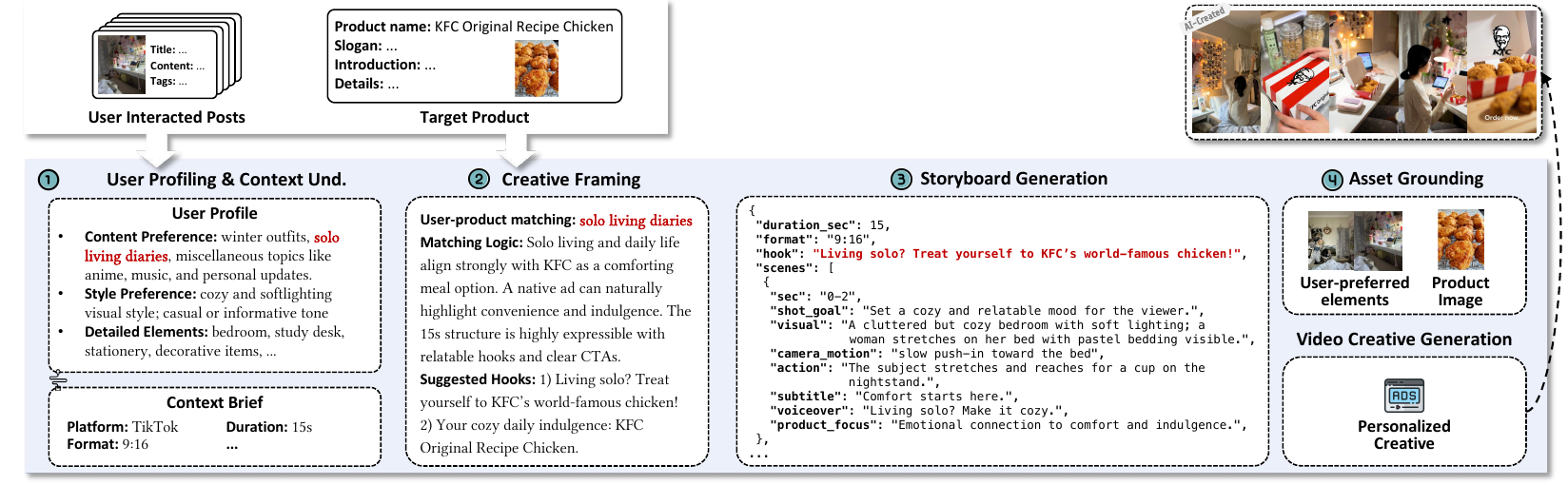}
\caption{\textbf{Overview of the PCG pipeline.}} 
\label{fig:pipe_pcg}
\end{figure*}

\begin{figure}[t]
    \centering
    \setlength{\abovecaptionskip}{0.2cm}
    \setlength{\belowcaptionskip}{-0.1cm}
    \includegraphics[width=\linewidth]{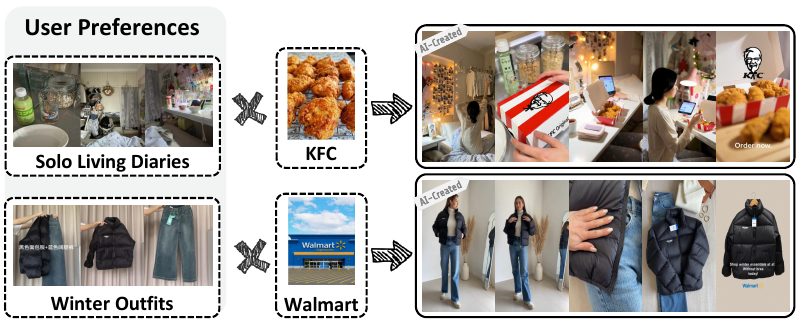}
    \caption{\textbf{Examples generated by NextAds on PCG-Bench.}}
    \label{fig:case_study1}
\end{figure}

\section{Disclosure Statement}
All product information and user profiles shown in figures are illustrative examples for research only, and do not imply endorsement, partnership, sponsorship, or any commercial affiliation. They must not be used for commercial advertising or promotion. If any IP, trademark, portrait rights concerns arise, the corresponding figures can be replaced with alternative illustrative assets.

\section{Implementation Details}

\subsection{Personalized Creative Generation}
\subsubsection{\textbf{Pipeline}}
\noindent For personalized creative generation, our pipeline consists of the following five key stages, as depicted in Fig.~\ref{fig:pipe_pcg}.
\begin{itemize}[leftmargin=*]
    \item \textbf{User profiling.} 
    To support precise user modeling, we build user profiles using a VLM to infer users’ textual and visual preferences from interaction history. We characterize textual preferences by topical interests and presentation style (\eg tone), and infer visual preferences by aggregating interacted images to identify preferred visual styles and elements.
    
    \item \textbf{Creative framing.} Next, conditioned on the user profile and target product information, we employ the VLM to evaluate product–interest compatibility across the user’s multiple topical interests and select the best-matching topic/scenario, thereby naturally connecting user interests to commercial intent.

    \item \textbf{Storyboard generation.} Based on the user profile and matching result, the VLM generates an executable, scene-level storyboard script, which specifies camera motions, narration pacing, and visual composition for every 2--3 second slot.
    
    \item \textbf{Asset grounding.} To support faithful generation and mitigate hallucinations, we construct a reference visual collage that combines user-preferred visual elements with official product images. This collage serves as a strict grounding constraint, ensuring both personalization and alignment with the product.
    
    \item \textbf{Video creative generation.} Finally, given the storyboard and grounded assets, we use a video generation model to synthesize the final creative, blending the user-preferred topic and aesthetics with the product’s key features.
\end{itemize}

\begin{figure*}[t]
\setlength{\abovecaptionskip}{0.2cm}
\setlength{\belowcaptionskip}{-0.2cm}
\centering
\includegraphics[width=\linewidth]{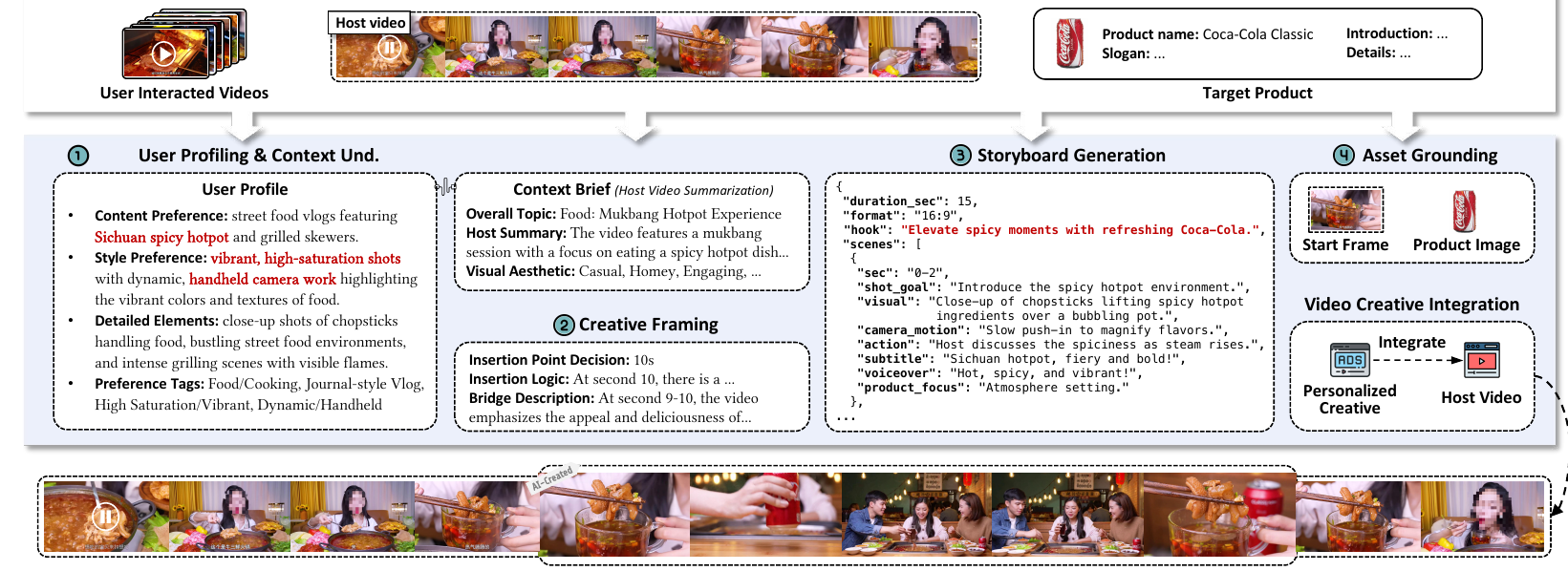}
\caption{\textbf{Overview of the PCI pipeline.}} 
\label{fig:pipe_pci}
\end{figure*}

\subsubsection{\textbf{Case Study}}
We present examples generated by closed-source NextAds in Fig.~\ref{fig:case_study1}. Based on user interaction history, the model accurately infers two dominant preferences (\ie \textit{Winter Outfits} and \textit{Solo Living Diaries}) with a stronger inclination toward the former. For Walmart, the model identifies high semantic compatibility between the user’s interest in \textit{Winter Outfits} and Walmart’s apparel-related offerings, and thus generates a hook centered on ``Winter fashion essentials under \$50'' to capture attention. In contrast, for KFC, the model pivots to \textit{Solo Living Diaries}, which better aligns with KFC’s themes of comfort and convenience, and constructs a storyline around ``Your cozy daily indulgence''. Overall, this adaptive user-product matching prioritizes user–product compatibility over naive top-interest matching, improving relevance while reducing the risk of semantic mismatch.

\subsection{Personalized Creative Integration}
\subsubsection{\textbf{Pipeline}}
For personalized creative integration, the pipeline consists of the following six key stages, as depicted in Fig.~\ref{fig:pipe_pci}.
\begin{itemize}[leftmargin=*]
\item \textbf{User profiling.}
Based on user interaction history, especially the videos they have engaged with according to the PCI-Bench, we use a VLM to preprocess these videos and annotate preliminary attributes, including topic, textual presentation style (\eg tone, narrative structure, and information density), and visual presentation style (\eg visual tone and camera motion). We then aggregate these annotations together with the corresponding video covers to construct a user profile that captures the user’s textual and visual preferences, as well as finer-grained preferred elements.

\item \textbf{Context understanding.}
We use a VLM to generate a structured summary of the host video, which captures four key aspects: topic, content, visual presentation, and audio characteristics. 

\item \textbf{Creative framing.}
Since the PCI task primarily targets seamless integration of the generated creative, we focus on integration smoothness and omit the user–product matching stage. Specifically, we employ a VLM to determine the optimal integration point. Given the target product information and the host video, the model identifies the frame that yields the most natural transition for introducing the product, and outputs 1) the chosen integration point with a brief rationale and 2) the corresponding start frame in the host video for integration.

\item \textbf{Storyboard generation.} 
Based on the user profile, target product information, host video summary, and the integration decision, the VLM generates a detailed storyboard script similar to the PCG pipeline. Importantly, the creative is required to start at the selected integration point and eventually return to it, ensuring a smooth transition back to the host video. This constraint encourages the generated content to blend naturally with the surrounding host video context.

\item \textbf{Asset grounding.}
Likewise, to promote faithful generation and reduce hallucinations, we build a reference visual collage by compositing the target product images, user-preferred elements, and the selected integration start frame, which provides explicit visual grounding that constrains the creative generation process.

\item \textbf{Video creative integration.}
Finally, we use a video generation model to synthesize the ad creative from the storyboard and grounded assets, and then insert the generated segment into the host video at the selected integration point.
\end{itemize}

\begin{figure}[t]
    \centering
    \setlength{\abovecaptionskip}{0.2cm}
    \setlength{\belowcaptionskip}{-0.2cm}
    \includegraphics[width=\linewidth]{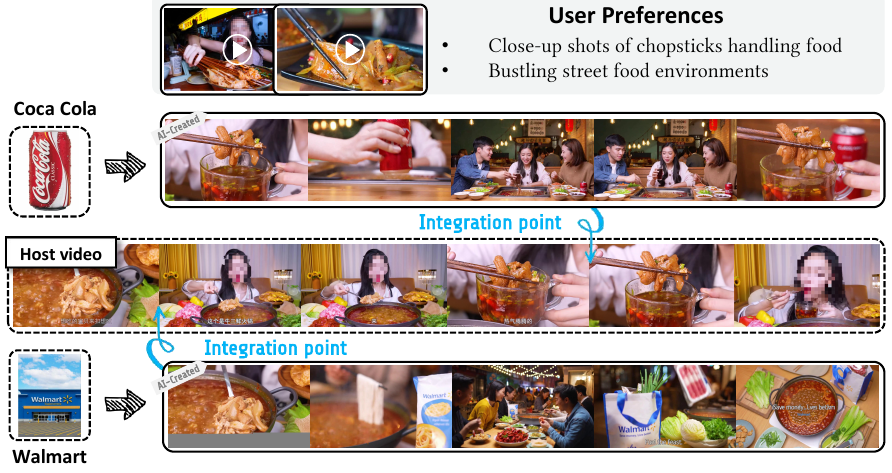}
    \caption{\textbf{Examples generated by NextAds on PCI-Bench.}}
    \label{fig:case_study2}
\end{figure}

\subsubsection{\textbf{Case Study}}
We present examples generated by the closed-source NextAds in Fig.~\ref{fig:case_study2}. Based on user interaction history, the model accurately infers user preference for Sichuan cuisine and lively, crowd-rich scenes, and incorporates these cues into the generated creatives. For Coca-Cola, the creative transitions smoothly from the host preparing to taste spicy food to picking up a bottle of Coca-Cola, positioning it as a complementary beverage to Sichuan hotpot. This preserves narrative continuity while grounding the product in the surrounding scene context. For Walmart, the host video segment showcasing hotpot ingredients naturally shifts to a Walmart supermarket scenario, where these ingredients are framed as being sourced from Walmart, enabling contextually coherent integration. Overall, these cases demonstrate that NextAds can effectively capture user preferences to generate personalized creatives for the target product, and seamlessly integrate the resulting ad segments into the host video.

\section{Evaluation Prompts}
\label{appendix:eval}

To assess the generated creatives, we adopt an LLM-as-a-judge setup and design five dedicated judge prompts, as follows.
\begin{center}
\begin{tcolorbox}[
                  breakable,
                  colbacktitle=gray!30,
                  coltitle=black,
                  colback=white,
                  colframe=black,
                  width=\linewidth,
                  arc=1mm, auto outer arc,
                  boxrule=0.5pt,
                  left=3pt,
                  right=3pt,
                  top=1pt,
                  bottom=1pt,
                  middle=1pt,
                  title={\textbf{Personalization -- Presentation}},
                  halign title=left,
                  toprule=0.5pt,
                  bottomrule=0.5pt,
                  before upper={\setlength{\parskip}{1pt}},
                  before lower={\setlength{\parskip}{1pt}},
                 ]
You are an expert in evaluating the presentation of creative content. Given a product and the covers of the user's interacted videos, evaluate how well the advertisement aligns with the user's aesthetic preferences regarding visual style. Consider the following factors:
\begin{itemize}[leftmargin=*]
    \item \textbf{Tone}: Does the tone match the user’s preference (warm, cold, neutral)?
    \item \textbf{Lighting}: Is the lighting appropriate for the user’s preference (soft, harsh, natural)?
    \item \textbf{Texture}: Does the texture (smooth, rough, \etc) fit with the user’s aesthetic?
    \item \textbf{Artistic style}: Does the artistic style (minimalistic, modern, vintage, \etc) align with the user’s taste?
\end{itemize}

\tcbline

\textbf{Criteria:}
\begin{itemize}[leftmargin=*]
    \item \textbf{Perfect match (10)}: The content presentation matches perfectly with the user's preferences.
    \item \textbf{Good match (7-9)}: The content presentation aligns well but has some minor differences.
    \item \textbf{Moderate match (4-6)}: The content presentation partially matches the user's preferences.
    \item \textbf{Poor match (1-3)}: The content presentation does not align well with the user's preferences.
    \item \textbf{No match (0)}: The content presentation completely fails to align with the user's preferences.
\end{itemize}

\tcbline

\textbf{Output JSON:}
\begin{lstlisting}[style=jsonstyle]
{
  "score": number,
  "reason": "brief analysis"
}
\end{lstlisting}
\end{tcolorbox}
\end{center}
\begin{center}
\begin{tcolorbox}[
                  breakable,
                  colbacktitle=gray!30,
                  coltitle=black,
                  colback=white,
                  colframe=black,
                  width=\linewidth,
                  arc=1mm, auto outer arc,
                  boxrule=0.5pt,
                  left=3pt,
                  right=3pt,
                  top=1pt,
                  bottom=1pt,
                  middle=1pt,
                  title={\textbf{Personalization -- Content}},
                  halign title=left,
                  toprule=0.5pt,
                  bottomrule=0.5pt,
                  before upper={\setlength{\parskip}{1pt}},
                  before lower={\setlength{\parskip}{1pt}},
                 ]
You are an expert in evaluating how well the advertisement matches the user’s interest. Given a product and the covers of the user's interacted videos, Evaluate how well the advertisement aligns with the user's aesthetic preferences regarding visual style. Consider the following factors:
\begin{itemize}[leftmargin=*]
    \item \textbf{Subject matter}: Does the subject matter of the advertisement align with the user’s preferences?
    \item \textbf{Scene semantics}: Does the scene selection (location, setting) align with what the user would enjoy?
    \item \textbf{Information angle}: Does the advertisement focus on the aspects the user cares about (performance, aesthetics, value for money)?
    \item \textbf{Appeal}: Does the ad focus on motivations that resonate with the user (saving money, efficiency, quality, aesthetics, health, eco-friendliness)?
\end{itemize}

\tcbline

\textbf{Criteria:}
\begin{itemize}[leftmargin=*]
    \item \textbf{Perfect match (10)}: The content is perfectly aligned with the user's preferences.
    \item \textbf{Good match (7-9)}: The content aligns well but with some differences.
    \item \textbf{Moderate match (4-6)}: The content partially aligns with the user's preferences.
    \item \textbf{Poor match (1-3)}: The content does not align well with the user's preferences.
    \item \textbf{No match (0)}: The content is completely irrelevant to the user.
\end{itemize}

\tcbline

\textbf{Output JSON:}
\begin{lstlisting}[style=jsonstyle]
{   
    "score": number, 
    "reason": "brief analysis"
}
\end{lstlisting}

\end{tcolorbox}
\end{center}
\begin{center}
\begin{tcolorbox}[
                  breakable,
                  colbacktitle=gray!30,
                  coltitle=black,
                  colback=white,
                  colframe=black,
                  width=\linewidth,
                  arc=1mm, auto outer arc,
                  boxrule=0.5pt,
                  left=3pt,
                  right=3pt,
                  top=1pt,
                  bottom=1pt,
                  middle=1pt,
                  title={\textbf{Product Alignment}},
                  halign title=left,
                  toprule=0.5pt,
                  bottomrule=0.5pt,
                  before upper={\setlength{\parskip}{1pt}},
                  before lower={\setlength{\parskip}{1pt}},
                 ]
You are an expert at evaluating the consistency of products in the advertisement. Given a product and the video, Evaluate how well the advertisement matches the product based on the following factors:
\begin{itemize}[leftmargin=*]
    \item \textbf{Product identity accuracy}: Does the ad represent the product’s appearance, color, model/version, packaging, and accessories correctly?
    \item \textbf{Brand and recognition elements}: Does the ad correctly display the product’s logo, branding, and design language?
    \item \textbf{Selling points and fact consistency}: Does the ad’s selling point match the product’s real features and details?

\end{itemize}

\tcbline

\textbf{Criteria:}
\begin{itemize}[leftmargin=*]
    \item \textbf{Perfect match (10)}: The product in the video matches the actual product in every way.
    \item \textbf{Good match (7-9)}: The product is mostly consistent with minor details changed.
    \item \textbf{Moderate match (4-6)}: The product is partially consistent, but with some significant differences.
    \item \textbf{Poor match (1-3)}: The product in the video is mostly inconsistent with the actual product.
    \item \textbf{No match (0)}: The product identity is completely inconsistent with the actual product.
\end{itemize}

\tcbline

\textbf{Output JSON:}
\begin{lstlisting}[style=jsonstyle]
{   
    "score": number, 
    "reason": "brief analysis"
}
\end{lstlisting}

\end{tcolorbox}
\end{center}
\begin{center}
\begin{tcolorbox}[
                  breakable,
                  colbacktitle=gray!30,
                  coltitle=black,
                  colback=white,
                  colframe=black,
                  width=\linewidth,
                  arc=1mm, auto outer arc,
                  boxrule=0.5pt,
                  left=3pt,
                  right=3pt,
                  top=1pt,
                  bottom=1pt,
                  middle=1pt,
                  title={\textbf{Integration Quality}},
                  halign title=left,
                  toprule=0.5pt,
                  bottomrule=0.5pt,
                  before upper={\setlength{\parskip}{1pt}},
                  before lower={\setlength{\parskip}{1pt}},
                 ]
You are an expert in evaluating the seamlessness and intrusiveness of advertisement within video content. Given a video containing an advertisement, evaluate how naturally the ad is integrated or how disruptive (intrusive) it feels to the viewer's experience. Consider the following factors:
\begin{itemize}[leftmargin=*]
    \item \textbf{Contextual flow}: Does the transition between the organic content and the advertisement feel smooth, or is it a sudden, jarring break?
    \item \textbf{Thematic consistency}: Does the ad share a similar tone, visual style, or topic with the surrounding video content?
    \item \textbf{User experience disruption}: Does the ad interrupt a high-stakes moment (e.g., a climax or a key explanation), or is it placed at a logical breaking point?
    \item \textbf{Narrative integration}: Is the ad woven into the story/dialogue (native advertising), or is it an external overlay/hard cut that pulls the viewer out of the experience?

\end{itemize}

\tcbline

\textbf{Criteria:}
\begin{itemize}[leftmargin=*]
    \item \textbf{Seamless (10)}: The ad is perfectly integrated; the viewer might not even perceive it as a disruption.
    \item \textbf{Natural (7-9)}: The ad is well-placed and shares thematic elements with the video, causing minimal friction.
    \item \textbf{Noticeable but Acceptable (4-6)}: The ad is clearly a break from content, but occurs at a logical pause or maintains a similar aesthetic.
    \item \textbf{Intrusive (1-3)}: The ad feels forced, breaks the immersion significantly, or uses a jarringly different tone/volume.
    \item \textbf{Highly Disruptive (0)}: The ad completely ruins the viewing experience through poor timing, extreme contrast, or aggressive interruption.

\end{itemize}

\tcbline

\textbf{Output JSON:}
\begin{lstlisting}[style=jsonstyle]
{   
    "score": number, 
    "reason": "brief analysis"
}
\end{lstlisting}

\end{tcolorbox}
\end{center}
\begin{center}
\begin{tcolorbox}[
                  breakable,
                  colbacktitle=gray!30,
                  coltitle=black,
                  colback=white,
                  colframe=black,
                  width=\linewidth,
                  arc=1mm, auto outer arc,
                  boxrule=0.5pt,
                  left=3pt,
                  right=3pt,
                  top=1pt,
                  bottom=1pt,
                  middle=1pt,
                  title={\textbf{Diversity}},
                  halign title=left,
                  toprule=0.5pt,
                  bottomrule=0.5pt,
                  before upper={\setlength{\parskip}{1pt}},
                  before lower={\setlength{\parskip}{1pt}},
                 ]
You are an expert at evaluating the diversity of content. Evaluate the diversity between advertisements for the same user on different products, considering the following aspects:
\begin{itemize}[leftmargin=*]
    \item \textbf{Visual style}: Is there excessive similarity in tone, lighting, texture, and artistic style across different product ads?
    \item \textbf{Narrative form}: Is the narrative structure (tutorial, vlog, plot twist, etc.) and tone (professional, humorous, etc.) overly similar across products?
    \item \textbf{Subject matter and scene semantics}: Is the subject matter or scene selection overly similar across ads targeting the same user?
\end{itemize}

\tcbline

\textbf{Criteria:}
\begin{itemize}[leftmargin=*]
    \item \textbf{10}: Ads are completely distinct with no noticeable overlap.
    \item \textbf{7–9}: Ads are mostly distinct with only minor similarities.
    \item \textbf{4–6}: Ads are somewhat similar but still distinct in key areas.
    \item \textbf{1–3}: Ads are largely similar with only slight variations.
    \item \textbf{0}: Ads are identical in all aspects.

\end{itemize}

\tcbline

\textbf{Output JSON:}
\begin{lstlisting}[style=jsonstyle]
{   
    "score": number, 
    "reason": "brief analysis"
}
\end{lstlisting}

\end{tcolorbox}
\end{center}

\end{document}